\documentclass[preprint,12pt]{elsarticle}
\usepackage{amssymb}
\usepackage{color}
 \newcommand{\comments}[1]{}   
\journal{Comptes Rendus de l'Acad\'emie de Sciences}

\begin{document}
 
\begin{frontmatter}

\title{Out of equilibrium dynamics of classical and quantum complex systems}
 \author{Leticia F. Cugliandolo} 
\address{Universit\'e Pierre et Marie Curie
  - Paris 6, Laboratoire de Physique Th\'eorique et Hautes Energies,
  4, Place Jussieu, Tour 13, 5\`eme \'etage, 75252 Paris Cedex 05,  France}
 
\begin{abstract}
Equilibrium is a rather ideal situation, the exception rather than the rule in Nature. Whenever the external or internal parameters 
of a physical system are varied its subsequent relaxation to equilibrium may be either impossible or take very long times. 
From the point of view of 
fundamental physics no generic principle such as the ones of thermodynamics allows us to fully understand their 
behaviour. The alternative is to treat each case separately.  
It is illusionary to attempt to give, at least at this stage, a complete description of 
all non-equilibrium situations. Still, one can 
try to identify and characterise some concrete but still general features of a class of out of equilibrium problems - yet to be identified - 
and search for a unified description of these. In this report I briefly describe the behaviour and theory of a set of 
non-equilibrium systems and I try to highlight common features and some general laws that have emerged 
in recent years.
\end{abstract}


\end{frontmatter}


\tableofcontents

\section{Introduction}
\label{sec:introduction}

This text is a stroll in the non-equilibrium 
dynamics of  classical and quantum physical systems with a few or many degrees of freedom,
closed or in contact with an environment. 
While in classical mechanics the theory of dynamical systems has dealt with very rich out of equilibrium 
phenomena for more than a century, in condensed matter or statistical physics focus was set 
on equilibrium properties until relatively recently. Microcanonical, canonical or grand canonical ensembles were used depending on the 
conditions of preference or relevance. The relaxation of a tiny perturbation away from equilibrium is also sometimes 
described in textbooks and undergraduate courses. However, important problems in physics and other sciences have 
obliged us to move away from the equilibrium hypothesis and face the difficulty of grasping and explaining 
out of equilibrium dynamics in these areas as well. 

The scope of this report is a rather wide one and, in consequence, the presentation is not intended to be 
technical. I will shortly describe various out of equilibrium situations in different fields of 
science ranging from physics to biology and including present-day themes of study in computer science. 
Beyond understanding each of these and other problem on their own, the 
ultimate aim of the analysis should be to identify generic features 
(dynamic scaling, thermodynamic-like properties, {\it etc.}) and methods of study
that could be of wide applicability.  Success in this direction has been achieved in recent years and 
I will shortly describe some of these remarkable findings.

As departure from equilibrium can be found at all scales, some of the 
examples are formulated in terms of variables that behave classically
but others, drawn from the microscopic world, are set in terms of quantum mechanics.
In the instances that I have chosen to treat relativistic effects are not important and will 
be neglected throughout. I organise the body of the text in three main sections, dealing with background, 
classical dynamics and 
quantum problems. For lack of space, and also because many of the questions posed apply to both 
realms albeit in a slightly modified way, I devote a more extended presentation to classical problems.
In the concluding Section I list a number of open problem that I 
personally find interesting and at analytic reach.

\section{Background}
\label{sec:classical}

I start by recalling the basics of
classical mechanics, the construction of equilibrium statistical mechanics, the emergence of 
collective phenomena
and phase transitions, and some peculiar effects induced by quenched disorder, competing 
interactions, and constraints.  I next explain why these systems give rise to yet not well 
understood, nor satisfactorily described, evolutions.

\subsection{Dynamical systems}
\label{subsec:dynamical-systems}

\textcolor{black}{\it Dynamical systems}~\cite{dynamical-systems} consist of classical variables that,
given an initial condition,  
evolve in time following some deterministic rule that univocally yields the trajectory 
in phase space. The evolution laws 
can be discrete in time, and be called maps, or they can be continuous. Maps, and in particular cellular automata~\cite{Droz}
in which the dynamic variables take only a discrete set of values, are useful to  model non-physical 
systems. Population dynamics in which each time step represents
a generation are an example. In the microscopic modelling of physical systems time is a continuous variable.
The classification and understanding of all possible dynamic trajectories
for different choices of the rule, embedding box and initial conditions 
is the goal of  this branch of theoretical physics. A \textcolor{black}{\it conservative} dynamical system conserves the volume in
phase space while a \textcolor{black}{\it dissipative} one does not.
A specially interesting question
is whether qualitative changes affect the trajectories 
when some parameter is varied at, {\it e.g.}, {\it bifurcation points}. The theory of \textcolor{black}{\it deterministic chaos} deals with 
non-linear systems with a few degrees of freedom that, in spite of following a deterministic dynamic rule, show a sustained erratic 
temporal behaviour. 
Chaotic  systems are extremely sensitive to small perturbations on the initial conditions.
Coherent collective behaviour in \textcolor{black}{\it coupled map lattices}~\cite{coupled-maps},
such as global synchronisation and spatio-temporal patterns, 
occur for strong spatial coupling,  while 
the propagation of a perturbation in both space and time
when a large number of individual chaotic maps are set in interaction
has been observed  in the weak coupling regime. 
Such persistent disorder in space and time
called \textcolor{black}{\it spatio-temporal chaos} is caused by instabilities in the 
deterministic dynamics and not by external noise. A traditional example
with these phenomena is  the atmosphere-ocean evolution.  
Unless otherwise stated, these problems feel no influence from the environment.

In \textcolor{black}{\it Hamiltonian dynamical systems} the evolution is
dictated by Hamilton's equations of motion that do conserve the volume in phase space. Whenever the Hamiltonian
does not depend explicitly on time the mechanical energy is a constant of the motion.
The simplest example is the one of a single classical particle embedded  in a $d$ dimensional space with some volume, 
possibly taken to be infinite. 
Hamiltonian dynamical systems are said to be \textcolor{black}{\it integrable} when there exist as many (linearly independent)
constants of the motion as, say, coordinate type degrees of freedom.

The theory of dynamical systems has reached maturity many years ago and, after having recalled some definitions 
and introduced some terminology, I will not discuss it further here. Excellent
books describe it from the very mathematical to the rather practical viewpoints~\cite{dynamical-systems}. 


\subsection{Equilibrium}

The \textcolor{black}{\it ergodic hypothesis}
states that the diverging time average along the trajectory of every observable expressed as an integrable function,
and for almost all initial conditions, equals 
its average over a measure on phase space that is left invariant by the dynamics. This hypothesis is the usual starting
point for the development of \textcolor{black}{\it equilibrium statistical mechanics}, 
the central postulate of  which
states that, after all transient effects have died out, an isolated Hamiltonian system can be found  in each of its accessible states
with equal probability. This principle introduces time-independent probabilistic concepts in the description of dynamical systems
and establishes the \textcolor{black}{\it microcanonic} distribution. 

Questions on ergodicity and equilibration of
closed systems have recently regained interest, boosted by the development of quantum systems in almost
perfect isolation. We will come back to this problem in Sec.~\ref{sec:quantum}.
However, these issues are highly non-trivial  in the classical limit as well.
Integrable systems are not ergodic. Non-integrability, as introduced by non-linear terms, 
does not necessarily imply the rapid establishment of ergodicity, as demonstrated by the 
almost exact  finite-period recurrence of the 
celebrated Fermi-Pasta-Ulam model of a large number of one dimensional linear harmonic oscillators 
anharmonically coupled~\cite{Fermi-Pasta-Ulam}. 
What initially appeared as a paradox was later shown to be a fact, with the Kolmogorov-Arnold-Moser
theorem that asserts that the evolution of a large set of initial conditions of an integrable
 Hamiltonian system non-linearly perturbed can remain quasi-periodic if the strength of the 
 perturbation is sufficiently small (how small is small is a hard problem-dependent question).

Accepting that ergodicity established, one rarely uses the microcanonic  ensemble.
The  \textcolor{black}{\it canonical} (or grand-canonical) equilibrium distribution is built by partitioning the very large system into a 
subsystem of interest and a much larger part that is effectively `integrated out'. Through some mathematical manipulations 
(that need the additivity of the  energy of the two subparts and the large size limit of the `bath'-part to 
ensure the absence of temperature fluctuations)
one derives the  equilibrium distributions of energy (and particle number) fluctuations of the selected part. 
The connection with macroscopic observables and thermodynamics is next established. 

The same procedure can be formulated dynamically, starting from the Hamiltonian evolution 
of the full coupled system~\cite{Weiss99}. With convenient
choices for the bath model and the coupling between bath and system the calculations can be carried out 
to derive effective dynamic equations for the `reduced system'. Under further assumptions on the initial conditions
of the bath degrees of freedom that typically introduce stochasticity into the description 
one deduces \textcolor{black}{\it generalised 
Langevin equations} for the selected part. This calculation gives a formal  foundation to one of the 
most commonly used tools to tackle the dynamics of open classical systems.


\subsection{Macroscopic systems: phase transitions}

One of the most striking features of large equilibrium systems is their collective behaviour and
the possibility of undergoing \textcolor{black}{\it phase transitions}, that is to say,
sharp changes in the macroscopic behaviour at special values of the parameters~\cite{phase-transitions}. 
Phase transitions can usually be detected with the measurement of an order parameter,  
generically defined as the average of a simple observable ({\it e.g.} the magnetisation density in a magnet) 
that vanishes in one phase and is different from zero in another. 
They can only arise
in the limit of a diverging number of degrees of freedom as it is associated to the non-analyticity
of the free-energy. 
In the Ginzburg-Landau approach one postulates the order-parameter(s) dependent 
free-energy the extrema of which characterise the equilibrium and metastable states in the different 
phases. This idea is at the basis of the \textcolor{black}{\it free-energy landscape} widely used in the description
of disordered systems that we will sketch below and in Sec.~\ref{subsubsec:structural}.

In \textcolor{black}{\it first order phase transitions} the order parameter jumps at
the critical point  to a finite value right on the ordered side of the transition. 
This is accompanied by discontinuities in 
various thermodynamic quantities and the divergence
of a first derivative of the free-energy density. 
The high and low temperature phases coexist at  the transition. These transitions often exhibit hysteresis and
memory effects since the system can remain in the metastable phase
when the external parameters go beyond the critical point. Examples are 
the melting of $3d$ solids and the condensation of 
a gas into a liquid. 

In a \textcolor{black}{\it second order phase transition} there is no phase coexistence  since 
the phases on either side of the transition are identical at the critical point.  The order parameter is continuous 
but the linear susceptibility, a second derivative of the free-energy, diverges.
By virtue of the fluctuation-dissipation theorem this
 implies the divergence of the correlation length of the order parameter fluctuations and the scale-free character of the  
critical state. These transitions can be classified in \textcolor{black}{\it universality classes} depending on 
space dimensionality, the dimension of the order parameter, and the symmetries of the problem. 
They are usually accompanied by \textcolor{black}{\it spontaneous symmetry breaking},
{\it i.e.}, the fact that the system has several equivalent equilibrium states related by symmetry in the 
ordered phase. 
\textcolor{black}{\it Spontaneous broken ergodicity} is 
another common feature of these transitions. Once the system is ordered in one of the components  of the equilibrium distribution, 
the long-time average (in which the system remains in one the disjoint ergodic components)
differs from the statistical one (performed over all components).
One can reconcile the two results by, in the statistical average,
summing over the configurations in one of the ergodic components only or, in the dynamic approach, taking the 
diverging time limit before the infinite size one. 

The Kosterlitz-Thouless transition is special in that it separates a disordered
from a critical phase. Topological defects proliferate in the disordered
phase and they bind in pairs in the one with quasi long-range order and correlation functions that 
decay algebraically. Physical realisations are
$2d$ planar ferromagnets, superconducting films, Josephson-junction
arrays, especially tailored nematic liquid crystals and toy models for two-dimensional
turbulence. The $2d$ xy model is the paradigm in this class of \textcolor{black}{\it topological phase transitions}.

All these features are, of course, very well established. The general picture is captured by mean-field theory
while the critical properties are obtained with perturbative renormalisation group (RG) techniques~\cite{phase-transitions}.

\subsection{Some interesting problems}

We introduce now a number of problems in which the picture sketched above 
gets more complicated, already at a static level.

\subsubsection{Disorder}
\label{subsubsection:disorder}

No material is perfectly homogeneous: impurities of different kinds are distributed
randomly throughout the samples. Such \textcolor{black}{\it disorder} can be of two types:
annealed or quenched. In the former case, the relaxation times of the impurities and host variables are 
 of the same order and the equilibrium
properties of the full system are given by the 
partition sum over all configurations of host 
and impurities. In the latter, there is a sharp separation of time-scales: the one for the  
impurities is much longer than the one of the host variables. 
It is the case of some magnetic systems  in which the diffusion time of the impurities 
is so long that these remain frozen.

A natural effect of disorder on many body systems is to favour disordered configurations ({\it e.g.}, to lower the critical 
temperature). However, 
as quenched randomness naturally generates a complex potential energy landscape
with many wells and barriers that scale and are organised in a highly non trivial way,
novel phenomena are also encountered.

As a particular sample is not expected to be 
special in any sense, the full ensemble of statistically equivalent samples 
is studied at once, assuming that for most interesting observables the typical 
behaviour is equal to the mean. 
A certain number of specific prescriptions to carry out the 
average  exist: 
the replica theory, the supersymmetric method, and the dynamical approach~\cite{Jorge}. 
Effective interactions are introduced by the 
averaging procedure but no spatial heterogeneities are left over. Scaling or RG methods to be mentioned
below allow one to 
make concrete predictions on the effect of spatial fluctuations induced by disorder.

Much attention has been payed to the effect of \textcolor{black}{\it weak quenched disorder} (inducing no frustration)
on systems in which the nature of the phases are not
modified by the impurities but the critical phenomenon is (see, {\it e.g.}~\cite{Berche} and refs. therein). On the one hand, the
critical exponents of second order phase transitions might be modified by disorder;
on the other hand, disorder may smooth out the discontinuities of first order phase
transitions rendering them continuous or even pushing them to zero temperature, 
as it was rigorously proven to occur in the 
$2d$ random field Ising model. 
The Harris criterion estimates the fluctuations of a second order 
critical temperature induced by the disorder spatial fluctuations. 
Finite regions of the system can order due to fluctuations in the quenched randomness
 below the critical point of the pure case and above the one the disordered case.
These properties manifest in non-analyticities of the free-energy, 
the so-called Griffiths singularities. For instance, deviations from Curie-Weiss ($\chi = 1/T$) behaviour 
appear below the N\'eel temperature of dilute antiferromagnets in a uniform field. 

\textcolor{black}{\it Spin-glass materials} are the archetype of systems with \textcolor{black}{\it strong disorder}. 
They are very imbalanced mixtures of a majority non-magnetic (sometimes metallic) 
with a minority magnetic element. The compound is prepared in the liquid phase and
it is very rapidly cooled down and let solidify. The location of the magnetic impurities is random and 
the ensuing ferromagnetic and antiferromagnetic interactions induce frustration.
These systems show a second-order phase transition between a paramagnetic 
and a spin-glass phase in which the spins take local and 
different preferred directions. In the SK  model, defined on a complete graph with two-spin interaction
strengths drawn from a probability distribution function, the spin-glass phase is 
very peculiar, with a very complex 
free-energy landscape as function of local magnetisations. 
The order parameter is no longer a simple observable 
but a relatively complex functional.
These results,  found with the replica theory~\cite{spin-glass-beyond},  were
later confirmed by the analysis of metastable states and, more recently, pure probability 
theory~\cite{Talagrand}.

The study of the ordered static phases in systems with strong quenched 
randomness beyond mean-field is very difficult. The nature of the spin-glass phase has been a matter of debate
for more than 30 years as the competing droplet scaling picture~\cite{droplet} assumes that it  is just a ferromagnet
in disguise with two equilibrium states related by spin reversal symmetry. 
Which of the two pictures  prevails in finite dimensions remains an open question.

Very few analytic techniques have been successfully applied to finite dimensional strongly disordered models
in equilibrium. Four kinds of RG methods are currently being explored.
The first one is a real-space
 variation of the Migdal-Kadanoff block renormalisation on regular lattices. Differently from this, the  Dasgupta-Ma 
 or strong disorder RG 
consists in the progressive elimination of  the
degrees of freedom with higher energy, to obtain an effective low-energy theory.
 It assumes that disorder-induced fluctuations dominate with respect to 
 any other source of fluctuations. It becomes asymptotically exact if the distribution of disorder broadens 
 without limit at large scales~\cite{Igloi-Monthus}. The method has been successfully applied to random quantum spin chains, 
the random field Ising model, low-dimensional diffusion in random media, notably, the 
Sinai problem. The field theoretical functional RG
takes into account the evolution of the statistic properties of the 
full random potential under re-scaling. 
This method has been mainly used to study  interfaces in random media but connections with the SK 
model have also been established~\cite{LeDoussal}. Finally, extensions of the 
so-called exact RG to deal with the effects of disorder are currently being investigated~\cite{Delamotte}.

\subsubsection{Structural glasses}
\label{subsubsec:structural}

In \textcolor{black}{\it structural glassy systems}  the dynamic variables are not quenched but still interact 
in a competing way. Their low-temperature configurations may look disordered but still have 
macroscopic properties similar to those of a crystalline state. Structurally a liquid and a  glass
look similar but while the former cannot support stress and flows, 
the glass has solid-like properties as crystals, 
it supports stress, and does not easily flow in 
reasonable time-scales.
A dynamic crossover between these two `states' is only observed. 

The importance of the glassy problem is stressed by the ubiquitous existence of materials with the above-mentioned properties.
Glasses  occur over an astounding range of scales.
Macroscopic examples include granular media like sand and powders. Unless 
fluidized by shaking or during flow these quickly settle into jammed, amorphous configurations. Jamming can also be caused by applying 
stress, in response to which the material may effectively convert from a fluid to a solid, refusing further flow. 
Colloidal suspensions 
contain smaller (typically micrometre-sized) particles suspended in a liquid and form the basis of many paints and coatings and at high 
density tend to become glassy unless crystallisation is specifically encouraged. 
On smaller scales there are atomic and molecular glasses such as  window glass 
formed by quick cooling of a silica melt or the polymer plastics in drink bottles. 
Finally, on the nanoscale, vortex lines in type-II superconductors also form glasses. 
Lennard-Jones binary mixtures undergo a glassy crossover when cooled across a temperature $T_g$ or when compressed across a density $n_g$. 

In mean-field  disordered systems such as, for instance, a directed elastic manifold 
embedded in an infinite dimensional space under a random potential with short-range correlations, 
or  a spin model on a fully connected graph with multi-spin interactions,  a 
mixed kind of phase transition occurs in which the order parameter is discontinuous 
but all first derivatives of the free-energy density remain finite. 
A more careful analysis of the free-energy density as a function of local order parameters
indicates the following organisation~\cite{Binder-Kob,Berthier-Biroli,Cugliandolo}.
A  high-temperature disordered phase is associated to a single absolute minimum of it. 
At a temperature commonly called $T_d$ 
the Gibbs-Boltzmann measure is broken into an exponentially large (in the number of degrees of freedom) 
number of pure states (local minima of the order-parameter-dependent free-energy).
Averaged observables are the result of the weighted average over these states and, surprisingly enough,
a smooth continuation of the high temperature behaviour is found and the thermodynamic free-energy remains analytic. 
The entropy of these
states (defined as the logarithm of its number) is called \textcolor{black}{\it complexity}, and it 
vanishes (\textcolor{black}{\it entropy crisis}) at a static transition temperature, $T_s<T_d$. 
For $T < T_s$ the states that yield the dominant
contribution to the statics are the equilibrium glassy phase. 
This is called a \textcolor{black}{\it random first order transition}.
Although these models have quenched
random variables, differently from structural glasses that do not, their behaviour
resembles in so many ways the one of \textcolor{black}{\it fragile glasses} that they are accepted 
(at least by a part of the community) as their mean-field model.
In finite dimensions transitions should be replaced by crossovers and infinite life-times 
by very long ones.

\subsubsection{Constraints and frustration}

Frustration is the name given to  the impossibility
of satisfying all competing interactions simultaneously.  
In certain systems, the local minimisation of the interaction energy on a frustrated unit gives rise to a 
macroscopic degeneracy of the ground state for some sets of parameters, 
unconventional phase transitions and the emergence of critical 
phases among other interesting features. Traditional  examples are 
\textcolor{black}{\it constrained classical magnets}, with anti-ferromagnets on a triangular lattice 
as the most prominent one.

Similar features are observed in systems with 
hard local constraints that are the result of complex microscopic interactions.
Spin-ice samples are materials of this kind~\cite{constrained-magnets}.
These systems exist naturally in $3d$ and have been engineered in the 
laboratory in $2d$. On the theoretical side, the celebrated \textcolor{black}{\it vertex models} are 
a quite incredible playground that realises many of these special features. 

\subsubsection{Physics and computer science}
\label{subsubsec:computer}

\textcolor{black}{\it Combinatorial optimisation}  
can be formulated as the minimisation of a cost or energy function defined over  
elementary degrees of freedom, that can be recast as one instance of 
a spin model on a random graph and with disordered many-body interactions~\cite{Mezard}.
The randomness in the 
ensemble of instances is the exact parallel of the disorder in physical samples.  
The difference arises in the questions asked in one and the other context. 
While in the physical context ones determines averaged macroscopic 
quantities with the usual stat-mech approach, 
 in the optimisation context one seeks the microscopic configurations that are  the exact 
energy minima of, moreover, the \textcolor{black}{\it worst instance} of a problem.
Indeed, physics  focuses on averaged properties and usually disregards worst case events. 
With this {\it proviso} in mind, physical tools have still been
useful to better understand optimisation 
problems.

 Analytic tools borrowed or inspired from disordered spin models such as the 
cavity approach and belief or survey propagation methods~\cite{Mezard}
gave a rather complete understanding of the statistical properties of these problems.
A phase transition between a phase with
one or many zero energy states and one in which the absolute energy minimum is strictly 
positive has been identified.
Interesting optimisation problems, for instance, constrained satisfaction ones, turn out to have a cost 
function landscape similar to the one of disordered spin models of the random first order transition class, with a 
partition of the solution space into clusters with a finite configurational entropy and a condensation transition associated to this complexity 
tending to zero. This complex structure influences the performance of physical algorithms. 

Algorithms that use the physically gained knowledge of the free-energy landscape in phase space have proven useful to 
tackle \textcolor{black}{\it typical} randomly generated instances of hard optimisation problems. The hope is that 
knowledge on the averaged and typical behaviour might be helpful to design algorithms to attack {\it hard} single instances as well.
For instance, it was proposed to tune a 
simulated annealing procedure to slowly follow thermal configurations towards a ground (zero-energy) state until the 
condensation transition but not beyond. The threshold of algorithms that do not respect local physical 
laws may be better but it will depend on the stochastic rules  used to wander in configuration space.

\section{Classical dynamics}

I here discuss, in some detail, the dynamics of classical complex problems.

\subsection{Dynamics close to equilibrium}

Onsager theory (reciprocity relations) 
characterises the behaviour of observables that are conserved in the full isolated systems
({\it e.g.} energy or matter), when these are perturbed not very far from equilibrium with a variation of their 
associated intensive parameter ({\it e.g.} temperature or pressure). A linear Taylor expansion relates the current of the 
extensive quantities and the variation of their 
associated intensive parameter {\it via} kinetic or transport coefficients ({\it e.g.}, the familiar heat equation).
This approach naturally leads to an exponential decay of the perturbation towards equilibrium. 
(Exponential relaxation near equilibrium is not necessarily enforced when one adopts
a more microscopic approach.)

The stochastic dynamics of systems perturbed away from equilibrium by an external
perturbation or an internal fluctuation are also well documented in the 
literature. Equilibration has consequences
upon (many) time-dependent correlation functions: they are stationary, meaning that 
they just depend upon the dynamics during the time delays involved in their definition but not upon the starting 
time of the measurement,  and they are linked to linear responses (susceptibilities)
by fluctuation-dissipation theorems~\cite{Cugliandolo}. 

We will not expand upon the dynamics close to equilibrium here as our
interest is in far from equilibrium situations. We just wanted to mention 
these two well-known facts to go beyond them below.

\subsection{Dynamics far from equilibrium}

The main reasons for being far from equilibrium are the following. 

--
We have already mentioned the possibility that an isolated classical Hamiltonian 
system  of integrable or even non-integrable kind may not reach equilibrium
(KAM theorem). The fate of a quantum isolated system under similar conditions,
named a \textcolor{black}{\it quantum quench},
receives great attention nowadays. 

--
At the most extreme microscopic level a closed physical classical system should 
be described by an Hermitian Hamiltonian and its
dynamics be invariant under time-reversal.
In some cases (arguing some coarse-graining procedure) one departs from this natural rule
and ascribes {\it ad hoc} updates to the degrees of freedom 
(in continuous or discrete time) 
that may  not lead to equilibrium. 

--
The initial condition of an open system can be far away  from the 
equilibrium states at the working conditions and the \textcolor{black}{\it relaxation dynamics be too slow} 
to take the system from one to the other. In many 
cases of interest, the time needed to equilibrate is a growing function of the size of the confining box
 or the number of variables that may diverge for all practical purposes. This situation can be realised 
 classically and quantum mechanically as well.

-- 
Open systems can be maintained out of equilibrium  by external baths at different temperatures, 
particle reservoirs at different chemical potentials, 
external fields, 
{\it etc.} 
In such cases there is  transport of heat from the hot bath to the cold one, mass from one reservoir to another, {\it etc.}
Quite generally, a stationary state, in which instantaneous observables do not 
depend upon time,  establishes and it is called a  \textcolor{black}{\it non-equilibrium steady state (ness)}.
This setting also appears classically and quantum mechanically.

--
At  a coarse-grained level external drive on a classical stochastic system can be mimicked by
transition rules that do not satisfy detailed balance. Such dynamics may also lead to a {\it ness}. 

The above listed situations are realised in a number of problems that have, or still are, 
intensively studied. We discuss some of them   
and we simultaneously  sketch some methods used to deal with 
them below. 

\subsubsection{Dynamical systems}

We just mention here an idea developed in this field that has been recently 
borrowed to study the non-equilibrium dynamics of disordered systems. In the 70s
Ruelle proposed to construct a dynamic partition function
to count trajectories rather than microscopic states. This approach,
closely linked to \textcolor{black}{\it dynamic large deviation theory}, 
was worked out  in some simple problems
such as the  Lorentz lattice gas. A renewed interest in this approach 
was gained in recent years with the derivation of  \textcolor{black}{\it fluctuation theorems}~\cite{fluctuation-theorems},
see Sec.~\ref{subsec:fluctuation-theorems}, as a consequence of symmetry properties of such a dynamic large deviation function. 
The 
subsequent derivation of what is called \textcolor{black}{\it stochastic thermodynamics}~\cite{Sekimoto,Seifert} is also 
attracting much attention. The 
method is also used in the context of glassy systems (see, {\it e.g.},~\cite{Fred}).

\subsubsection{Open dissipative single particle systems}

Possibly, the simplest  open system with out of equilibrium dynamics -- though this is not 
always appreciated -- is \textcolor{black}{\it Brownian motion}, {\it {\it i.e.}} the erratic rattling 
of a colloidal particle immersed in a fluid and under no external force. 

Both experimentally and with, {\it e.g.},  the Langevin approach one observes that, for any set of initial conditions, 
after some relaxation time the velocity of the particle becomes Maxwell-distributed. This 
was the requirement used by Langevin to postulate the friction and noise term as well as its statistics. 
The velocity relaxation time depends upon the ratio between the 
particle mass and the strength of the friction 
and in most cases of interest is extremely short.
In these cases the inertia term can be dropped, and the Langevin equation 
becomes an over-damped first-order stochastic differential equation.
Among these instances are colloidal suspensions, a Ôsoft condensed matterÕ example; 
spins in ferromagnets coupled to lattice phonons, a Ôhard condensed matterÕ case; and 
proteins in the cell, a ÔbiophysicsÕ instance.
In contrast, the position does not reach an equilibrium measure. Indeed, the mean-square displacement 
increases  with time, {\it i.e.} the particle undergoes \textcolor{black}{\it diffusion}, unless the 
motion be confined by a finite box or an external potential. Moreover, position correlation 
functions \textcolor{black}{\it are not stationary} and the \textcolor{black}{\it fluctuation-dissipation relations} between 
its linear response and the correlation are not the ones dictated by equilibration~\cite{Cugliandolo}.

In this very simple example
one reckons that  different dynamic stochastic variables can behave very differently in the 
same time regime,   
with the velocity being equilibrated and the position undergoing out of equilibrium dynamics.

This is also the simplest problem in which quenched disorder plays a dominant role.
The diffusive properties of the particle can be heavily 
influenced by a quenched random potential in ways that depend upon its
statistical properties (range of correlations) 
and the dimensionality of space~\cite{Bouchaud-Georges}. 
The problem has applications to the hopping conductivity of disordered materials and 
the diffusion of a test particle in a porous medium among many others.
It has also been used as a caricature of the dynamics of many-body complex
systems with glassy nature, in which the particle represents the 
system's configuration and its motion occurs in a configurational space 
decorated with a disordered  landscape. 
A weak effect of quenched randomness is the modification of the transport 
coefficient, {\it e.g.}, the diffusion constant, while a much stronger one is the appearance 
of \textcolor{black}{\it anomalous diffusion}, {\it i.e.}, a different temporal dependence of the 
mean-square displacement, even for white noise. 
Perturbative~\cite{Bouchaud-Georges} and Dasgupta-Ma~\cite{Igloi-Monthus} RG techniques 
have been successfully applied to this problem. Anomalous diffusion has also been found in clean  but
geometrically constrained systems as, {\it e.g.}, on fractal structures, or in particles moving in flat regular spaces
but in contact with a bath with coloured noise. The latter problem is of biological interest, as 
one uses Brownian particles as tracers to characterise
the milieu in which they move.

\subsubsection{Elastic manifolds}

More interesting
features are expected to be found in the relaxation dynamics of many-body or extended systems.
The dynamics of 
manifolds with $d$ internal 
dimensions 
embedded in $N+d$ dimensional spaces with $N$ the dimension of the transverse space~\cite{Halpin-Healey,elastic-lines}
is often encountered in physics.
A directed line ($d=1$) mimics vortex lines in $N+d=3$ dimensional high-$T_c$ superconductors
or  stretched polymers. 
An interface is a frontier separating two regions covered by two phases. It could be the border between water and oil in a liquid mixture, 
or between positive and negative magnetisation in a magnet. 
Growth phenomena such as the burning front in a forrest, the advance of a crack in a rock, fluid 
invasion in porous media, the growth of a surface on a substrate due to material deposition combined (or not) with material diffusion, 
or even the growth of  a bacterial colony define interfaces that can be mimicked as elastic manifolds.
One distinguishes cases in which the interfaces are closed and therefore not directed, from those in which they are single-valued 
with respect to some reference plane and therefore directed. 
A manifold  in a random medium is sometimes considered to be a `baby' spin-glass problem due to the
frustration induced by the competition between the elastic energy that tends to reduce the deformations and quenched disorder that tends to distort the structure.


Interface models can be defined by restricted solid-on-solid models or with continuous equations.  
The former are discrete
and  advantageous to do numerical simulations.  Different rules for the particle deposition (random, ballistic, random with 
surface relaxation) are used.  
Continuous models often describe the surfaces at larger length-scales: a coarse-graining process is employed to describe the surface with a 
continuous function. Manifolds under the effect (or not) of quenched 
random potentials, with (Kardar-Parisi-Zhang) or without 
(Edwards-Wilkinson, Mullins-Herring) non-linear interactions, 
and with short- or long-range contributions to the elastic energy are modelled with variations of a  functional Langevin
equation. 

The morphology of an interface is usually characterised by its width.
Under certain circumstances interfaces \textcolor{black}{\it roughen}, that is to say,
the space-averaged dispersion of its position about 
its mean increases with time until reaching saturation 
at a value that grows with the linear size of the manifold. In the presence of quenched disorder the asymptotic 
roughness undergoes a crossover at a temperature
and disorder strength dependent scale. For interfaces with smaller linear length 
the statics and dynamics are controlled by thermal fluctuations. For interfaces with 
longer linear length disorder takes control and changes the scaling laws. These results were suggested 
by scaling arguments and they were later set into firmer grounds with the
functional RG~\cite{elastic-lines}. 

The relaxation dynamics of clean and disordered elastic manifolds, before the linear-length dependent 
crossover to saturation, occurs out of equilibrium and present 
 \textcolor{black}{\it ageing effects} and non-trivial linear responses, with \textcolor{black}{\it violations of the equilibrium 
 fluctuation-dissipation theorem}. Their behaviour has many points in common with the one
observed in glassy systems~\cite{Kolton09}.  A growing length can be identified in these problems
and its properties studied carefully.

The driven dynamics of an elastic directed  manifold in 
a random medium present two hallmarks  intensively studied in the last two decades. At
zero temperature a finite force  is needed to set the manifold into motion. The associated
\textcolor{black}{\it de-pinning transition} is a dynamic critical phenomenon with similarities with a second order 
phase transition, where the velocity plays the role of the order parameter and the force the one of the control 
parameter. At finite temperature the manifold moves for arbitrarily small forces
but the scaling of its mean velocity with force strength and temperature is highly non-trivial
characterising what is called \textcolor{black}{\it creep motion}~\cite{elastic-lines}. 

A ferromagnetic material, such as iron, under
a slow and smooth increase of a magnetic field gets magnetised through a random sequence of 
steps associated to changes in the size and orientation of  microscopic 
clusters of aligned atomic magnets (spins).  The magnetic changes are often detected {\it {\it via}} 
acoustic noise. The amount of such Barkhausen noise increases in systems with 
impurities, crystal dislocations, {\it etc.} and it is therefore used to detect the presence of unwanted internal 
material stresses and to test the sample's mechanical properties~\cite{Sethna}.
In general, the response of an elastic manifold to external driving through a 
disordered medium is not smooth, but
exhibits discontinuous and collective jumps called \textcolor{black}{\it avalanches}
which extend over a broad range of space and time scales.
The distribution functions of the bursts' magnitude and duration, as well as other characteristics, 
have been characterised. A self-similar pattern emerges with scaling laws~\cite{Pierre}.

The pulled averaged and fluctuating  dynamics of manifolds in random media have been
studied with the functional RG but the non-stationary relaxation in finite 
dimensional transverse space still  remains out of the reach of this technique. A \textcolor{black}{\it dynamic  Gaussian  
variational approach} in the limit of an infinite transverse space dimensionality is the sole analytic tool
we can count upon to study this question at present.
 
\subsubsection{Critical dissipative dynamics}

In \textcolor{black}{\it quenches to a critical point}~\cite{Corberi-Cugliandolo-Yoshino}, as realised by, {\it e.g.}, 
the instantaneous 
change of external temperature from, say, infinity to $T_c$, patches with equilibrium critical 
fluctuations grow in time but their linear extent never reaches the equilibrium correlation 
length that diverges. Clusters within clusters of neighbouring spins pointing in the same direction 
grow in time with fractal interfaces between them. The system 
builds correlated critical Fortuin-Kasteleyn clusters with fractal dimension 
$d_{\rm FK} = (d + 2 - \eta)/2$, with $\eta$ the 
anomalous exponent, in regions growing algebraically as $R_c(t) \simeq t^{1/z_{eq}}$, with 
$z_{eq}$ the dynamic critical exponent. 
The relaxation time diverges close to the phase transition as a power law of the distance to criticality
$\tau\simeq (g-g_c)^{-\nu z_{eq}}$ with $\nu$ the exponent that controls the divergence of the correlation length. 
Such a slow relaxation is named \textcolor{black}{\it critical slowing down}.
Scaling arguments and perturbative RG calculations~\cite{Hohenberg-Halperin}  give explicit expressions 
for many of the quantities of interest. Numerical simulations probe the exponents and scaling 
functions beyond the available perturbative orders. In the asymptotic time regime the 
space-time correlation functions have a \textcolor{black}{\it multiplicative}
scaling form as, for instance, $C(r,t) \simeq C_{\rm st}( r ) C_{\rm ag}(r/R_c(t))$ with 
$C_{\rm st}( r ) \simeq r^{-2(d-d_{\rm FK})} 
= r^{2-d-\eta}$ taking into account that the equilibrium structures have a fractal nature (hence their density 
decreases as their size grows) and the fact that the order parameter vanishes at the second order critical point. 
The dependence on $r/R_c(t)$ in the ageing factor expresses the similarity of configurations at different times once lengths 
are measured in units of $R_c(t)$. At distances and times such that $r/R_c(t) \ll 1$ the equilibrium power-law decay,
$C_{\rm st}( r )$, is recovered, thus $C_{\rm ag}(x) = 1$ at $x \to 0$. $C_{\rm ag}(x)$ falls off rapidly for $x\gg 1$ to ensure that 
spins be uncorrelated at distances larger than $R_c(t)$. Similar factorisations apply to higher order correlation functions
and linear responses. The equilibrium 
fluctuation-dissipation theorem holds at short time and length-scales, where the stationary relaxation
is relevant, but violations appear at long time and length-scales~\cite{critical-FDT}.

\subsubsection{Dynamics across a phase transition.}

When a system with a \textcolor{black}{\it first order phase transition} is taken to a region in the
phase diagram in which it is still locally stable but metastable with respect to the
new absolute minimum of the free-energy, its evolution towards the new equilibrium
state occurs by \textcolor{black}{\it nucleation} of the stable phase. The theory of simple nucleation~\cite{Binder} 
establishes that the time needed for one bubble of the stable state to conquer the
sample grows as an exponential of the free-energy barrier over the thermal energy available, 
$k_B T$. Once the bubble has
reached a critical size, that also depends on this free-energy barrier, it very rapidly
conquers the full sample. The textbook
example is the magnetic magnetisation reversal in, {\it e.g.}, an Ising model in equilibrium under a
magnetic field that is suddenly reversed. As multiple
nucleation and competition between different states intervenes the problem gets
harder to quantify and very relevant to the mean-field
theory of fragile structural glasses as realised by the random first order phase transition 
scenario.

Take a system with a well-understood \textcolor{black}{\it second order phase transition} across the critical point 
by tuning a control parameter from the disordered and symmetric 
to its ordered and symmetry-broken phase. The system will tend to order locally in one of the, possibly many, 
new stable states that have the same free-energy density.
\textcolor{black}{\it Phase-ordering}~\cite{Bray} is characterised by a 
patchwork of large domains the interior of which is basically thermalised in one of the 
equilibrium phases while their boundaries slowly move and tend to become smoother due to their elastic energy. 
The patterned structure is not quiescent, ordered regions tend to grow on average with a linear length 
$R(t)$,  but the time needed 
to fully order the sample diverges with the system size.
This picture suggests the splitting of the degrees of freedom (spins) into two categories, providing 
statistically independent contributions to correlation and linear response functions. 
Bulk spins contribute to a quasi-equilibrium stationary contribution, while 
interfacial spins account for the non-equilibrium part. 
At late times and in the scaling limit $r\gg \xi$, $R(t) \gg \xi$ and $r/R(t)$ arbitrary, with 
$\xi$ the equilibrium correlation length, 
the system enters a \textcolor{black}{\it dynamic scale invariant regime}
in which there exists a single characteristic length, 
 $R(t)$, such that the domain 
 structure is, in statistical sense, independent of time when lengths are scaled by $R(t)$. In practice, 
 this means that all
 time and space dependence in correlation functions appear as ratios between distances and $R(t)$. 
For instance, the space-time correlation decomposes as $C(r,t) \simeq C_{\rm st}( r ) + C_{\rm ag}(r/R(t))$, 
with $C_{\rm st}( 0 ) =1-\langle \phi\rangle^2$, $C_{\rm st}( r ) \to 0$ for $r\to\infty$, $C_{\rm ag}(0) = 
\langle \phi\rangle^2$ and $C_{\rm ag}( x ) \to 0$ for $x\to\infty$.
 In clean systems the characteristic length grows algebraically in time, 
 $R(t) \simeq t^{1/z}$ with $z$ a dynamic exponent that defines the dynamic universality class. 
 For systems with competing interactions this law can be notably slowed down and logarithmic 
 growth has been obtained in these cases.

Phase ordering kinetics are rather well understood qualitatively although a full 
quantitative description is hard to develop as the problem is set into the form of a non-linear
functional Langevin equation
for, say, a scalar  ($N=1$) field $\phi$ 
with a double-well Ginzburg-Landau free-energy 
and no small parameter. This problem is ubiquitous and hence very important. Coarsening 
has been identified in relatively complex systems such as frustrated magnets or kinetically 
constrained models (see Sec.~\ref{subsubsec:glassy-dynamics}). It is not clear yet whether 
all glassy dynamics can fall into some kind of coarsening albeit of yet unknown type.

\textcolor{black}{\it Weak quenched disorder}
renders the dynamics of macroscopic systems even slower than 
in clean cases. Take, for instance, of the $3d$ random field Ising model,  with  paramagnetic and ferromagnetic equilibrium 
phases, as an example.
For probabilistic reasons, the fields can be very strong and positive in some region of the sample and favour positive magnetisation,
and very strong and negative in a neighbouring region and favour negative magnetisation. It will then be very hard to 
displace the phase boundary and let one of the two states conquer the local volume. The probability of finding 
such rare regions can be quantified and the time needed to displace the domain wall can be estimated with the activated Arrhenius 
argument. They influence very strongly the equilibrium and out of equilibrium dynamics on both sides of  the critical point. The 
relaxation of a perturbation away from 
equilibrium becomes slower than exponential, the dynamic counterpart of the Griffiths essential 
singularities  of the free-energy. The assumption of a power law dependence of 
free-energy barriers with size combined with an Arrhenius argument suggests $R(t) \simeq \ln^{1/\psi} t$
for the growing length.
  
The dynamics of systems with \textcolor{black}{\it strong quenched randomness} is very complex. 
After a quench from high to  low temperatures they not only show very slow out of equilibrium
relaxation, never reaching thermal equilibrium, but  
they also display very intriguing memory effects under complicated paths in parameter space~\cite{Vincent}.
 Although it is not clear whether the dynamics occur {\it {\it via}} the growth of domains, scaling of dynamic correlation functions 
 describe numerical data quite precisely and, somehow surprisingly, with a power law $R(t)\simeq t^{1/z}$.
 This fact is not compatible with mean-field predictions of a much complex time-dependence
 that could perhaps only establish at much longer time scales. But the power-law growth of $R$ is not compatible with 
the droplet picture predictions either. The proposal in this model is that  droplet-like low-energy excitations of various 
sizes on top of the ground state should render the dynamics strongly heterogeneous both in space and time. 
At low enough temperatures the evolution should be dominated by thermal activation. The typical free-energy gap of a 
droplet with respect to the ground state and the free-energy barrier to nucleate a droplet are assumed to scale 
as $L^\theta$ and $L^\psi$, respectively, with $\theta$ and $\psi$ two 
non-trivial exponents. Static order is assumed to grow as in standard coarsening systems
with two equilibrium states related by symmetry.
Dynamical observables such as the two-time auto-correlation function 
should then follow universal scaling laws in terms of a growing length $R(t) \simeq \ln^{1/\psi} t$.
Before drawing conclusions one must keep in mind that the analysis of experimental and  numerical data is  difficult
given the limited range of time scales available in both cases. In~\cite{Corberi-Cugliandolo-Yoshino} an efficient strategy for data analysis 
with the goal of finding the best $R(t)$ for dynamic scaling is discussed and might be of help in future analysis of this question.

\subsubsection{Glassy dynamics}
\label{subsubsec:glassy-dynamics}

The \textcolor{black}{\it random first order scenario} of the glassy arrest developed from the exact solution to fully-connected
disordered spin models with thee (or more) body interactions (see Sec.~\ref{subsubsec:structural}). 
These models present a very peculiar
static and dynamic behaviour that is supposed to be the mean-field description of the glass phenomenology~\cite{Binder-Kob}.
The dynamic approach detects the change in free-energy landscape at $T_d$ since the relaxation time 
diverges at this temperature and the system cannot reach equilibrium for any temperature 
below it~\cite{Cugliandolo}. 
For the relaxation dynamics from a disordered initial configuration $T_s$ does
not play any role. The out of equilibrium dynamics are dominated by  threshold
states that are high in free-energy and have flat
directions in the free-energy landscape. The separation of time-scales observed in the 
relaxation of two-time correlation functions can then be rationalised as the sum of
two processes: a transverse and a longitudinal motion in phase space with the
former being similar to a confined relaxation while  the latter being associated to 
diffusion along flat channels. The latter shows ageing effects and scaling properties
although the putative growing length is not easy to visualise~\cite{Corberi-Cugliandolo-Yoshino}.

The equilibrium diverging relaxation time (of a local dynamics respecting detailed balance) close and above $T_d$
can be set in correspondence to the equilibrium diverging correlation length 
of a \textcolor{black}{\it point-to-set spatial correlation}~\cite{Guilhem}.
This is shown with a formal upper bound between relaxation time and correlation length
that applies to finite dimensional systems as well. 

Another approach to glassy dynamics is given by
\textcolor{black}{\it kinetically constrained models}~\cite{kinetically}. 
These are similar to lattice gases in that
one and only one particle can occupy each vertex of a lattice but 
only if a number of local conditions are met, typically, that there is at least a 
number of empty sites in the immediate neighbourhood. Single particle jumps between 
nearest neighbour sites under the same constraint provide the dynamics. The rates satisfy detailed balance with respect 
to a Boltzmann measure  factorized 
over different sites and  energetically trivial as the particles do not interact.
These rules are supposed to be the result of the coarse-graining of a dense molecular system, although there is no formal 
proof of this fact. 
These models are  trivial thermodynamically, but they may have a rich dynamics with ergodicity breaking, 
slow relaxation, ageing and many other interesting features for many choices of the constraints. 


\subsubsection{Externally driven systems}

The problems mentioned above can be subjected to external perturbations in 
the form of forces that do not derive from a potential, baths at different temperatures 
connected on the borders, {\it etc.} In this section I mention a number of  models that are 
held out of equilibrium by persistent driving forces 
that are receiving much attention at present.

\textcolor{black}{\it Driven lattice gases}~\cite{driven-diffusive}
of asymmetric exclusion type (ASEP)~\cite{Stinchcombe}
are used to mimic mass transport in various contexts. 
In $1d$ they model molecular motors motility along filaments in 
the cytoskeleton, vehicle circulation along highway lanes or generic 
queueing problems. 
These models consist of a linear lattice
with open or periodic boundary conditions. At each time step 
one particle is chosen at random and it moves to the next site in one 
direction with probability $p$ or backwards with probability $1-p$ only if that
site is empty. Jumps over particles are forbidden. 
In open cases particles can enter the lattice with a particular
probability and leave with another probability, typically at the ends of the line.  

The $1d$ ASEP~\cite{Stinchcombe} 
is quite special in that it is a genuine out of equilibrium model for 
which an exact solution for the stationary regime, with a current of particles, is known.
There is a huge mathematical and theoretical physics activity 
around these problems due to their connection with powerful techniques 
(matrix product Ansatz, the Bethe Ansatz, combinatorics, random matrix theory) 
and their mapping to other physical 
problems such as  interface growth and, in the particular, the KPZ equation
for long length and time scales, or the XXZ spin-$\frac{1}{2}$  quantum 
Heisenberg chain with appropriate boundary conditions.

\textcolor{black}{\it Reaction-diffusion processes}~\cite{Tauber05} involve 
(possibly different kinds of) particles that freely diffuse in space and 
undergo reactions (annihilation, coagulation or transformations from one type to another)
when they are within some prescribed range. 
Particle
propagation can be modelled as a continuous  or discrete time random walk, either on a
lattice or in the continuum. Such models have been used to model
population dynamics, epidemic spreading, chemical reactions, {\it etc.} 
Breaking  detailed balance allows for dynamic phase transitions 
in $1d$ or a non-vanishing dynamic order parameter in $2d$ 
models with continuous symmetry thus circumventing the Mermin-Wagner theorem.

These systems are quite special in that 
fluctuations govern their behaviour in a much more dramatic way than in equilibrium.
For instance, in cases in which spontaneous particle decay competes with 
a production or branching process, fluctuations may generate a phase transition and not only 
alter its properties. A mean-field treatment in which correlations and spatial fluctuations are 
neglected 
(kinetic rate equations) predicts a unique active phase while the careful treatment of spatial 
fluctuations
(with numerical simulations or approximate approaches) demonstrates
the existence of phase transitions between different asymptotic states of 
active (with a non-vanishing particle density) and inactive (just free of particles) kind. 
The continuous transition found 
 is analogous to a second-order equilibrium phase transition,
and requires the tuning of appropriate reaction rates as control parameters.

The analytic treatment of these problems is very rich.
The master equation for a reaction-diffusion process can be
re-expressed as a Schr\"odinger-like equation with non-Hermitian Hamiltonian
by using a second-quantized bosonic creation-annihilation operators 
(Doi-Peliti formalism). With the coherent state
representation and an approp riate continuum limit one
constructs a path-integral for a statistical field theory. At this point 
one can apply all the field-theoretical machinery and, in particular, 
develop RGs. 
In some cases the field theory can be 
transformed into a form that imposes an effective Langevin 
equation for a (complex) field with multiplicative non-equilibrium (imaginary) noise
that does not respect the Einstein relation between noise and dissipation. However,
in some cases 
which are the effective field theories  remains a hotly debated 
question and there is no complete phase transition classification 
for their non-equilibrium phase transitions.
Experiments on reaction-diffusion physical systems are still quite rare.

\textcolor{black}{\it Pattern formation}
is the spontaneous formation of macroscopic spatial structures in open systems constantly  driven 
far from equilibrium. Patterns can be  stationary in time and periodic in space, 
periodic in time and homogenous in space or periodic in both space and time.
The key words `dissipative structures' or 
`self-organisation' are also attached
to this phenomenon that was initially studied  
in fluid dynamics and chemical reactions but also appears in 
solid-state physics, soft condensed matter and nonlinear optics,
and is now central to biology with, for instance,
morphogenesis and the dynamics of active matter (see below) playing a predominant role. 
The traditional example is Rayleigh-B\'enard convection: a fluid placed between two infinite 
horizontal plates
that are perfect heat conductors at different temperature. At a threshold value of the 
temperature difference the uniform state with a linear temperature profile
becomes unstable towards a state with convective flow. In the context of chemistry the paradigm are systems with  competition between 
temporal growth rates and diffusivity of the different species.

The theory of pattern formation is mature and reviewed in 
great detail in~\cite{Cross-Hohenberg}. 
These systems are usually described with over-damped dissipative 
deterministic partial differential equations. The exact form of the 
equations depends on the problem at hand, 
ranging from Navier-Stokes equations 
for fluid dynamics to reaction-diffusion equations for chemical systems.
A well-studied case is the non-linear Schr\"odinger equation.

The spatio-temporal structures are found from the growth and saturation 
of modes that are unstable when a control parameter is increased beyond
threshold. A parallel between the kind of bifurcation of fixed-point solutions and the 
order of a phase transition can be established.
Concretely, a linear stability analysis of the uniform state
reveals the mechanism leading to the pattern formation. The analysis of
non-linear effects is 
often realised numerically.
Weak Gaussian noise can select a particular pattern among many equivalent ones.

The constituents of \textcolor{black}{\it active matter}~\cite{active-matter}, be them particles, lines or other, absorb energy 
from their environment or internal fuel tanks and use it to 
carry out motion. In this new type of soft condensed matter energy is partially transformed into 
mechanical work and partially dissipated in the 
form of heat. The units interact directly or through disturbances propagated in the medium. 
Some realisations are bacterial suspensions, the cytoskeleton in living cells, or 
even swarms of different animals. 

In other driven systems, such as sheared fluids, vibrated granular matter and driven vortex lattices, 
energy input occurs on the boundaries of the sample. Instead, in active matter 
energy input is located on the internal 
units and is homogeneously distributed in the sample. 
Moreover, the effect of the motors can be dictated by the 
state of the particle and/or its immediate 
neighbourhood and it is not necessarily fixed by an external field.

Active matter is typically kept in a non-equilibrium steady state
and displays out of 
equilibrium phase transitions that may be absent in their passive counterparts. 
It often exhibits unusual mechanical properties, very large responses to small 
perturbations, and large fluctuations 
not consistent with the central limit theorem. Much theoretical effort has been recently devoted to 
the description of different aspects, such as the self-organisation of living 
microorganisms, the identification and analysis of states with spatio-temporal structure, 
such as bundles, vortices and asters and the study of the rheological 
properties of active particle suspensions (to identify the mechanical 
consequences of biological activity). A rather 
surprisingly result was obtained with a variational solution to the 
many-body master equation of the motorised version of the hard 
sphere fluid often used to model colloids: instead of stirring and thus 
destabilising ordered structures, the motors do, in some circumstances, 
enlarge the range of stability of crystalline and amorphous structures relative to the ones with 
purely thermal motion. The relation of this fact with the emergence of a lower than ambient 
effective temperature (see Sec.~\ref{subsec:effec-therm}) was suggested.

\subsection{Effective thermodynamics?}
\label{subsec:effec-therm}

The quest for an approximate thermodynamic description of non-equili\-brium 
systems or, to start with, the identification of effective parameters acting as the equilibrium ones, 
has a long history that we will not review here. We will simply focus on one notion, the effective 
temperature, $T_{\rm eff}$, that has proven to be a successful concept in classical glassy physics. 
The answer to the simple-looking question `how does a glassy model respond to an 
infinitesimal perturbation?' had a perturbingly simple answer: the linear response is linked 
to the strength of the ambient noise and the spontaneous fluctuations through 
fluctuation-dissipation relations (FDRs) in a way that resembles strongly the equilibrium 
fluctuation-dissipation theorem (FDT)~\cite{Cugliandolo}. These relations were later 
shown to hold, at least within the time scales that a simulation can access, in a host of more 
realistic models including coarsening, driven powders, colloidal suspensions, 
models for high-$T_c$ superconductors, active matter and so on and so 
forth. The recognition that the FDRs could be grasped in terms of $T_{\rm eff}$ paved the way to a 
more intuitive interpretation of the dynamics of out of equilibrium systems with slow dynamics, and 
led to the extension of other thermodynamic concepts~\cite{Cugliandolo-Teff}. Within the RFOT 
description of fragile glasses $T_{\rm eff}$ is observable-independent, it can be deduced 
from the variation of the complexity (or configurational entropy) around the energy 
reached asymptotically (as in a micro-canonical definition) and it satisfies many other
 welcome properties. The $T_{\rm eff}$ idea found further support from 
the fact that it replaces the bath temperature in Fluctuation Theorems, see below,
when the system that is driven far from equilibrium is itself unable to equilibrate with its 
environment.  Some of these properties were also checked in a variety of glass models
 although in some of them, such as some kinetically constrained systems or trap models, 
 they do not hold. Despite an enormous effort, we still lack a satisfactory understanding of 
its microscopic origin and real-space interpretation as well as the precise limits of validity 
of its thermodynamic interpretation.

\subsection{Fluctuation theorems}
\label{subsec:fluctuation-theorems}

In the last decade a number of exact results for externally driven
systems were proven.
The fluctuation theorem characterises the fluctuations of the entropy production 
over long time-intervals in certain driven steady states. It was first observed 
numerically, and later proven for reversible hyperbolic dynamical systems and 
open systems with stochastic dynamics. This remarkable result suggested
the search of other relations of similar kind. 
Constraints on the probability distributions for work,
heat and entropy production, depending on the nature of the system and
the choice of non-equilibrium conditions, were found (Jarzinsky equation, 
Crooks relation, {\it etc.})~\cite{Seifert}. One especially appealing consequence of these relations is
that one could use them to obtain free-energy differences of equilibrium states
by performing out of equilibrium measurements.

 \section{Quantum physics}
 \label{sec:quantum}
 
 The study of canonical (or macro canonical) equilibrium properties 
of quantum strongly interacting many-body systems has been the main 
focus of condensed matter. These properties can be described with a 
 combination of mean-field techniques, RG and scaling arguments to the 
 same level of satisfaction as their classical counterparts.
  
 Having said this, an important difference with classical statistical physics is that the 
reduction of a coupled system into a subpart by the integration of 
a large piece that is taken to be the bath gives rise to long-range
interactions (in imaginary time) that have non-trivial effects on 
the behaviour of the selected part. Equilibrium phase transitions  can be modified by the bath both in location as in 
order. Quite generally, one observes that the order phase is enlarged by the 
coupling to the bath. In order to observe these effects it is necessary to use an 
exact treatment of the bath (and not a Markov approximation).

 Quantum non-equilibrium phenomena are continuously growing in importance but their 
 theoretical understanding is still at a very early stage. Significant advances in the field 
 of ultra cold atoms have allowed one to engineer quantum many-body systems in 
 almost perfect isolation from the environment~\cite{Bloch}. 
 This ensures their coherent unitary evolution for sufficiently long times to be 
 studied in detail. Moreover, 
 thanks to the ability to rapidly tune different parameters, {\it e.g.}, the 
 interaction strength between the atoms or the
 creation of controlled excitations, the non-equilibrium dynamics have been accessed. 
 
 Besides, the physical behaviour of traditional out of equilibrium dissipative quantum systems~\cite{Weiss99} is 
 also a topic of great interest. In condensed-matter systems one usually selects the relevant 
 variables, say the spins, and treats the coupling to all other variables as the coupling to the bath, 
 {\it e.g.}, phonons in a solid. In mesoscopic cases one is mostly interested in local couplings to 
 current-carrying leads, {\it i.e.} fermionic baths. In cold atomic gases the light fields confining the gas 
 can also lead to dissipative processes. 
 
 Therefore, 
 the problems discussed in Sec.~\ref{sec:classical} that were set in microscopic
terms naturally admit quantum extensions and physical realisations. In the 
following we briefly mention some problems and methods that are being used in this area 
at present.

The thermalisation (or not) of isolated quantum systems, evolving with unitary dynamics 
after an abrupt or slow change in a parameter,  
is actively being investigated~\cite{Polkovnikov,Dutta}. The analytic solution of some
models, especially integrable systems in  $1d$, has allowed one to discuss this problem in
great detail and suggested that generalised Gibbs ensembles could be the relevant 
measure to describe them~\cite{Rigol}. The possible connection with many-body localisation 
has also been evoked~\cite{Santoro}.
 Beyond integrability, numerical methods 
in the form of variations of the density matrix renormalisation group technique ($t$-DMRG) using matrix product 
states~\cite{Schollwoeck} allow for the almost exact simulation of large strongly 
interacting quantum systems in $1d$. The large $N$ quantum $O(N)$ model provides a 
natural playground for analytic calculations and $1/N$ expansions that suggest the existence of non-thermal 
long-lived quasi-stationary states in the symmetric phase~\cite{Berges} and other non-trivial effects. 
Approximate methods of analytic kind, as time-dependent variational approaches for strongly 
interacting fermonic systems ({\it e.g.} 
the Hubbard model)~\cite{Schiro,Gabi} or numerical type, as 
diagrammatic Monte Carlo techniques and others~\cite{Pollet} are currently being used in this 
context. Tools  developed by the glassy systems community, as the use of 
fluctuation-dissipation relations (see Sec.~\ref{subsec:effec-therm}) of quantum nature in this case,  
may be useful to investigate whether usual thermal 
equilibration establishes (or not) in this context~\cite{Foini}.

We have already mentioned the non-trivial effect of a quantum bath on the phase transitions 
of large quantum system. A sufficiently strong quantum bath may also affect the evolution of 
a single quantum degree of freedom as shown by the highly non-trivial localisation transition of a two-level
system under the effect of a bath made of quantum oscillators at zero temperature~\cite{Leggett}.
A complete treatment of the memory induced by the bath is necessary to capture this 
phenomenon as well.

Impurity motion in low-$d$ quantum liquids~\cite{Cazalilla} has been a major field of research in the 
last decade, now realised in cold atom experiments.  The `artificial' experimental design of 
these systems has now become possible by confining cold atoms in optical 
nanotubes. Using these techniques, the diffusion of impurity atoms in contact with a Luttinger liquid 
(LL) with tuneable impurity-LL interaction was studied. Due to the external trapping potential the 
minority atoms undergo damped oscillations indicating that dissipation takes place in this system. 
This problem is one realisation of \textcolor{black}{\it quantum Brownian motion} in a confining
potential~\cite{Weiss99,Grabert}. It can be dealt with with analytic methods, e.g. quantum generating functionals, 
but also with the $t$-DMRG where the 
bath is taken to be constituted by a finite number of atoms confined in the same harmonic trap.

A \textcolor{black}{\it quantum phase transition} occurs at zero temperature and it is driven uniquely by quantum fluctuations. 
 It corresponds to a change in the symmetry of the ground state when the strength of the quantum
 fluctuations are appropriately tuned at zero temperature.
 Experiments on atoms trapped in an optical lattice that undergo a Mott insulator transition
 have prompted the study of equilibrium behaviour and 
 non-equilibrium dynamics close to  and across quantum phase transitions~\cite{Bloch}. 

\textcolor{black}{\it Quantum coarsening phenomena} and \textcolor{black}{\it quantum critical dynamics} have 
not been studied in sufficient depth yet. One case that has been studied in detail is 
the mean-field representation as realised by, {\it e.g.}, the large $N$ quantum $O(N)$ model coupled to 
an ensemble of quantum harmonic oscillators or to electron baths, has been 
studied in detail~\cite{Camille}. A separation of time-scales, as in the classical limit, and 
dynamic scaling with the classical growing length were found. The short-time and short-length 
dynamics are controlled by quantum and thermal fluctuations in equilibrium while the long
scales behave as classically with ageing and breakdown of the quantum fluctuation-dissipation 
theorem that takes, in this regime, a classical form. The non-trivial effect of the quantum bath was also reckoned. 
However, other relatively simple problems such as  
dissipative quantum spin chains set out of equilibrium remain to be analysed at the same level of completeness. 
(Some attempts to extend $t$-DMRG methods to include the coupling to a bath 
may be of help in this context, see, {\it e.g.},~\cite{Corinna}.)

It is well-known that quenched disorder can have fantastic effects on quantum systems
with Anderson localisation as the most prominent example. I will not dwell upon this 
or other well-understood problems such as transport in mesoscopic systems here. 
Concerning the physical systems discussed in Sec.~\ref{sec:classical}, 
one of the hallmarks of finite dimensional quantum 
spin models with weak disorder are Griffiths-McCoy singularities of their free-energy, 
that lead to a highly non trivial paramagnetic phase and critical behaviour~\cite{Vojta}.
Spin-glass phases have been identified in many condensed
matter systems at very low temperature. \textcolor{black}{\it Quantum glassy phases} exist also in 
electronic systems and structural glasses.  Wigner glasses are other examples of 
relevance modelled as elastic systems under the effect of quenched randomness.

The mean-field analysis of quantum dissipative systems with quenched randomness
can be performed with the same level of detail as the one for the classical 
counterparts. A definition and study of metastable states 
(in phase space) in mean-field quantum glasses 
has been done with an extension of the Thouless-Anderson-Palmer approach. 
The results are compatible with  the analysis of the free-energy density with the replica Matsubara approach. 
The out of equilibrium relaxation of these systems coupled to a quantum environment, for a generic initial density 
matrix not correlated with the quenched randomness hence mimicking a quench 
from the disordered phase, can be investigated with the Schwinger-Keldysh closed-time formalism. 
For details on these calculations see~\cite{Cugliandolo-Manybody} and references therein. Even more 
so than in the classical case, going beyond mean-field is extremely hard.

In spin models on hyper-random graphs with finite or infinite connectivity
and multi-spin interactions the random first order phase transition
becomes a genuine first order one at low 
temperatures~\cite{Cugliandolo-Manybody}.
This fact has formidable consequences to the failure of quantum annealing 
methods to solve hard optimisation problems that can be mapped, 
as already mentioned in Sec.~\ref{subsubsec:computer}, onto these models.
 The quantum annealing or adiabatic quantum algorithm is the quantum counterpart of thermal simulated
 annealing. The idea is to change the parameters in a quantum Hamiltonian to take the system 
 across a convenient path in the phase diagram. The progressive vanishing of quantum fluctuations 
should then  allow it to go from a simple ground state to the one of the hard classical problem one is actually interested in.
 This strategy will obviously fail when going across a first order transition line~\cite{Bapst}.

A problem of great interest is the characterisation of the ({\it e.g.}, mass or energy)
current flowing through a system coupled to external reservoirs at different 
temperatures or chemical potentials on its borders. Again, the literature on this 
topic is huge. I just want to mention here that 
the \textcolor{black}{\it full counting statistics}, that is to say,
all cumulants of the current have recently been computed with 
conformal field theory methods in $1d$ quantum problems at criticality~\cite{Doyon}
(see this reference for other methods applied to these same problems).

By the end of our discussion on classical systems I mentioned two 
generic ideas that have attracted considerable attention in recent years:
effective temperatures in systems in slow relaxation or weakly driven 
by external forces (see Sec.~\ref{subsec:effec-therm}) and fluctuation theorems (see Sec.~\ref{subsec:fluctuation-theorems}). Deviations from the equilibrium 
fluctuation-dis\-si\-pa\-tion theorem in quantum (mean-field) systems were 
found~\cite{Cugliandolo-Teff} and more 
recently some driven quantum problems were analysed from 
this perspective as well (see, {\it e.g.}~\cite{Caso}). However, the thermodynamic
sense of this notion in the quantum context has not been explored yet. 
In parallel, quantum fluctuation theorems have been proposed and are
currently being explored. I cite here only one very recent proposal to analyse 
quantum work statistics with an experimental set-up~\cite{Dorner} from 
which the rich literature on this subject can be re-constructed.

\section{Conclusions}

In the body of these notes I have tried to summarise our current 
understanding of out of equilibrium complex systems by focusing on a certain 
number of problems. Non-equilibrium physics is a very wide area of 
research and it is also rapidly evolving. My summary is necessary
partial. In particular, I have tried to cite review articles that could give a
global description of the topics touched upon in the text. Consequently, 
I have not made justice to original articles in the field.

In this Section I will simply list a number of problems that, in my view, would be 
interesting, and possible, to address theoretically in the near future.

-- Conformal field theory combined with Coulomb gas methods yielded a 
myriad of results on a large 
diversity of equilibrium geometric objects at $2d$ critical points
(critical percolation, self-avoiding walks, loop-erased random walks, q-states Potts models, 
{\it etc.}). More recently, conformally invariant stochastic growth, the so-called SLE~\cite{Cardy05}, 
allowed one to re-derive these results in more mathematical terms as well as to extend them in 
various directions. Could these be of help in out of equilibrium cases as well? 

--
Phase ordering kinetics or coarsening is a qualitatively well-understood phenomenon~\cite{Bray}. 
However, there is no performant analytic tool allowing to prove dynamic scaling, derive the 
growing length systematically, obtain the scaling functions, {\it etc.} This remains an open 
problem since long ago.

--
I gave a short description of the statics and dynamics of models with quenched 
disorder that is based on the exact solution to mean-field models~\cite{Binder-Kob,Berthier-Biroli,Cugliandolo}. Whether these
results hold true in finite dimensions is still an open question. The improvement of various 
RG techniques may be of help to decide upon this point.

-- The same {\it incognita} concerns the behaviour of glassy systems with no quenched 
randomness~\cite{Binder-Kob}.

As already said, most of these questions can be posed including quantum fluctuations.

\vspace{1cm}

\noindent
{\bf Acknowledgements:}
I wish to thank all my collaborators in this field for many years of very interesting exchanges and 
ANR-BLAN-0346 (FAMOUS) for financial support.

\end{document}